\documentclass[journal]{IEEEtran}

\usepackage{cite}

\usepackage[pdftex]{graphicx}
\graphicspath{{./figures/}}
\DeclareGraphicsExtensions{.pdf,.jpeg,.png,.jpg,.JPG}
\usepackage{caption} 
\usepackage{adjustbox}

\usepackage[cmex10]{amsmath}
\DeclareMathOperator*{\argmax}{arg\,max}
\usepackage{bbold}
\usepackage[oldsyntax]{stackengine}
\newcommand\shifthat[2]{%
  \stackengine{\Sstackgap}{$#2$}{\(\hspace{#1}\hat{}\)}{O}{l}{F}{T}{S}}
\newcommand\newhat[1]{%
\if A#1\shifthat{5.2pt}{#1}\else
\if B#1\shifthat{4.8pt}{#1}\else
\if x#1\shifthat{3.6pt}{#1}\else
\shifthat{3.4pt}{#1}
\fi
\fi
\fi
}

\usepackage{algorithm}
\usepackage{algpseudocode}

\usepackage{array}

\ifCLASSOPTIONcompsoc
 \usepackage[caption=false,font=normalsize,labelfont=sf,textfont=sf]{subfig}
\else
 \usepackage[caption=false,font=footnotesize]{subfig}
\fi

\usepackage[section]{placeins}
\usepackage{url}
\pdfoutput=1
\begin{document}
\title{Multi-Target Multiple Instance Learning for Hyperspectral Target Detection}

\author{Susan~Meerdink*,
        James~Bocinsky*,~\IEEEmembership{Student Member,~IEEE}, 
        Alina~Zare,~\IEEEmembership{Senior Member,~IEEE},
        Nicholas Kroeger,
        Connor~McCurley,~\IEEEmembership{Student Member,~IEEE}, 
        Daniel~Shats, 
        Paul~Gader,~\IEEEmembership{Fellow,~IEEE}
\thanks{*Authors have equal contribution to manuscript. Authors are affiliated with the University of Florida. S. Meerdink$^{1,2}$, J. Bockinsky$^{1}$, A. Zare$^{1}$, N. Kroeger$^{3}$, C. McCurley$^{1}$, D. Shats$^{1}$, and P. Gader$^{2,3}$ }
\thanks{$^1$ Department of Electrical and Computer Engineering} \thanks{$^2$ Engineering School of Sustainable Infrastructure and Environment} \thanks{$^{3}$ Department of Computer and Information Science Engineering} \thanks{(e-mail: susanmeerdink@ufl.edu; azare@ece.ufl.edu)
}
}

\markboth{}%
{Shell \MakeLowercase{\textit{et al.}}: Bare Demo of IEEEtran.cls for Journals}

\maketitle

\begin{abstract}
In remote sensing, it is often challenging to acquire or collect a large dataset that is accurately labeled. This difficulty is usually due to several issues, including but not limited to the study site's spatial area and accessibility, errors in the global positioning system (GPS), and mixed pixels caused by an image's spatial resolution. We propose an approach, with two variations, that estimates multiple target signatures from training samples with imprecise labels: Multi-Target Multiple Instance Adaptive Cosine Estimator (Multi-Target MI-ACE) and Multi-Target Multiple Instance Spectral Match Filter (Multi-Target MI-SMF). The proposed methods address the problems above by directly considering the multiple-instance, imprecisely labeled dataset. They learn a dictionary of target signatures that optimizes detection against a background using the Adaptive Cosine Estimator (ACE) and Spectral Match Filter (SMF). Experiments were conducted to test the proposed algorithms using a simulated hyperspectral dataset, the MUUFL Gulfport hyperspectral dataset collected over the University of Southern Mississippi-Gulfpark Campus, and the AVIRIS hyperspectral dataset collected over Santa Barbara County, California. Both simulated and real hyperspectral target detection experiments show the proposed algorithms are effective at learning target signatures and performing target detection.  
\end{abstract}

\begin{IEEEkeywords}
target detection, target characterization, hyperspectral, adaptive cosine estimator, spectral matched filter, multiple instance, multiple target
\end{IEEEkeywords}

%
\IEEEpeerreviewmaketitle

\section{Introduction}
\IEEEPARstart{H}yperspectral data is well suited for discrimination between a target and background because the hundreds of narrow bands can be used to identify subtle spectral shifts caused by materials’ differences in chemistry, physiology, and structure. However, each pixel measures the interaction of electromagnetic radiation with multiple surface constituents, regardless of spatial resolution \cite{Keshava2002}. Often targets of interest do not comprise a whole pixel resulting in a mixed-signal and require sub-pixel target detection. However, nearly all hyperspectral target detectors rely on having an accurate target spectral signature in advance. Usually, target signatures are obtained from spectral libraries collected in either controlled laboratory settings, outdoor hand-held spectrometer measurements, or pulled manually from a hyperspectral image. Spectra collected in controlled laboratory settings often do not match atmospheric or lighting conditions present in the hyperspectral imagery. Outdoor hand-held spectrometer measurements can have similar issues if not collected during hyperspectral image collection. Both of these methods require knowing and measuring the majority of materials present in the imagery. 

Often pulling spectra manually from a hyperspectral image is ideal because spectra capture the current environmental and lighting conditions. However, this method of developing spectral libraries can be complicated from many sources but can be mostly summarized into three categories. First, precise training labels for targets are often difficult to obtain for multiple scenarios. For example, a target's GPS coordinates can have errors of several meters or pixels. As a result, the target's real location might be several meters from the measured coordinates. Additionally, `pure' target pixels, or pixels that contain 100\% of the target, are challenging to find on the landscape. Secondly, the number of training pixels for a target class is small compared to the non-target training pixels. Especially for hyperspectral images covering a large spatial area, there are often only a few pixels for target training. Lastly, because each pixel is an interaction of multiple surface constituents, many targets are sub-pixel, and the targets' proportion is unknown. These complications can make obtaining the best-suited target signature difficult, which will ultimately drive the success of the target detection algorithm. Spectral libraries developed using outdoor hand-held spectrometer would also inherently contain these complications. 

\begin{figure}[!h]
    \centering
    \includegraphics[width=3.45in]{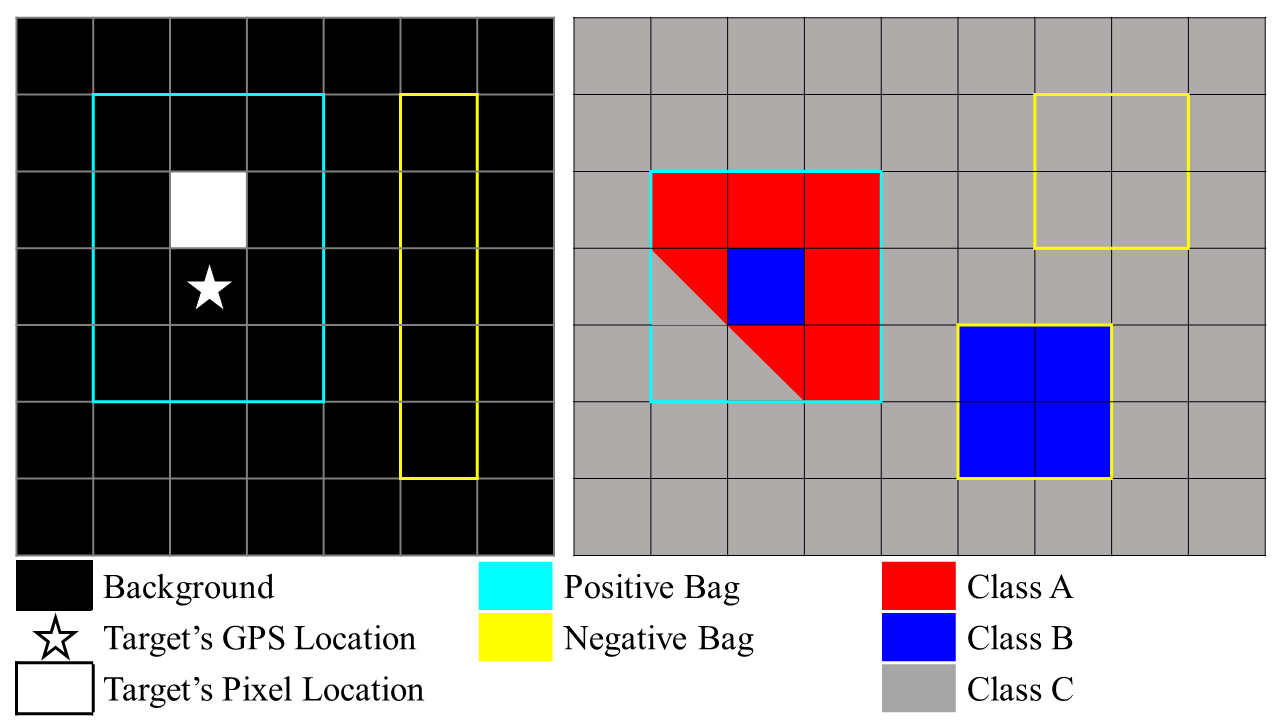}
    \caption{Simplified positive and negative bagging examples with each square representing a pixel in an image. The right figure shows when a target’s GPS location does not match the target’s pixel location in the imagery. This example represents the bagging methodology used on the MUUFL Gulfport dataset. The left figure shows when reference polygons are used for a class and may not have all pixels in a polygon representing the same class. This example describes the bagging methodology used on the AVIRIS Santa Barbara dataset.}
    \label{fig_bagexample}
\end{figure}

Multiple instance learning (MIL) can overcome the need to have precise training labels \cite{dietterich1997solving, Maron1998}. MIL only requires the labeling of positive and negative bags, which are groupings of pixels. Each bag may contain many pixels, but a bag is labeled positive if at least one of the pixels in it falls within the target class (Figure \ref{fig_bagexample}). This framework alleviates the need to have accurate labels which are inherently challenging to collect. Since the introduction of MIL, numerous MIL algorithms have been proposed \cite{zhong2019multiple, zare2018discriminative, jiao2018multiple, jiao2017multiple}. A pair of MIL algorithms known as the multiple instance adaptive cosine estimator (MI-ACE) and the multiple instance spectral match filter (MI-SMF) have been used with hyperspectral data and have shown competitive results with other algorithms in terms of single target concept estimation and detection \cite{zare2018discriminative}. However, these algorithms only determine one signature for a target. Most targets contain enough within-class spectral variability that it is difficult to capture that variability with a single signature. For example, if the target is a tree species, then the spectral signatures can vary significantly between individuals due to differences in structure, biochemistry, or phenology. An urban target can contain a lot of spectral variability due to material diversity (e.g., concrete, asphalt, paint) that also changes with age. 

In this paper, we propose the Multi-Target Multiple Instance Adaptive Cosine/Coherence Estimator (MTMI-ACE) and Multi-Target Multiple Instance Spectral Match Filter (MTMI-SMF) algorithms to extend the MIL framework and learn multiple target signatures compared to a single target signature. The objective of the MTMI-ACE and MTMI-SMF algorithms is to learn a dictionary of target representations, focusing on maximizing the detection of those targets against a background. Our overarching aim is to demonstrate the improvements and advantages of these algorithms for hyperspectral target detection.

\vspace{-5mm}

\subsection{Related Work}

Two of the earliest proposed MIL algorithms are the Diverse Density algorithm \cite{maron1998framework} and the Expectation-Maximization with the Diverse Density (EM-DD) algorithm \cite{ZhangGoldman2002}. These algorithms learn a target concept that is close to the intersection of as many positive bag instances while being as far from any negative bag instances. These algorithms learn a single target signature and use Euclidean distance to measure the similarity between instances. These two methods introduced the noisy-OR model that many later MIL algorithms relied upon \cite{srinivas1993generalization}. Although foundational, these two methods were not designed with hyperspectral detection in mind, and other methods soon outperformed these for hyperspectral applications. 

Another family of algorithms that belong to the MIL framework are the Functions of Multiple Instances (FUMI) algorithms. The original FUMI algorithm \cite{zare2010pattern} extends the approach of the Sparsity Promoting Iterated Constrained Endmember (SPICE) algorithm \cite{zare2007sparsity}. SPICE is an unsupervised algorithm that learns the sub-pixel proportions and endmembers of an unlabeled dataset for unmixing imagery. FUMI extends the SPICE algorithm by using labeled data to learn a target's signature as well as non-target endmembers. Variations of FUMI have been developed, such as convex FUMI (cFUMI) \cite{jiao2015functions} and extended FUMI (eFUMI) \cite{jiao2015functions}. While cFUMI assumes the exact target locations in a training image are known, the eFUMI algorithm needs only an approximate knowledge of target locations in training data. eFUMI learns multiple target signatures by determining the convex combinations of target and non-target signatures using an expectation-maximization approach. Through this approach, the algorithm focuses on learning the discriminative features between different target types, resulting in greater target characterization and discrimination. However, eFUMI estimates signatures and does not discriminate prototypes. Additionally, the number of targets and background signatures are needed to find target signatures which requires domain knowledge of the dataset. 

The MI-ACE algorithm estimates a single target signature that optimizes the widely-used ACE sub-pixel target detector on a training dataset with multiple-instance-style imprecise labels. The MI-SMF algorithm does the same but using the SMF sub-pixel target detector. However, these algorithms assume that a target's spectral variability can be captured with a single target signature. 

Another recent MIL algorithm is the Multiple Instance Hybrid Estimator (MI-HE) algorithm. The MI-HE algorithm learns multiple targets and background signatures that maximize the probability that positive bags are labeled positive, and negative bags are labeled negative \cite{jiao2018multiple}. The objective function is simplified by only needing to maximize a single instance from each positive bag. MI-HE makes the noisy-OR model more flexible by implementing a hyperparameter adjustable generalized mean to vary the operation between a min and max operation. Additionally, the algorithm solves for a sparsity vector to support dictionary element diversity. The MI-HE algorithm also determines multiple target signatures by using a data mixing model and optimizing the response of the hybrid sub-pixel detector within a MIL framework. The algorithm iterates between estimating a set of representative target and non-target signatures and solving a sparse unmixing problem. Although this algorithm performs highly competitively with other MIL algorithms, the algorithm has a complex number of hyperparameters and takes significantly longer to train than other MIL algorithms.

The most recent algorithm added to the MIL framework and hyperspectral target detection is the Multiple Instance Learning for Multiple Diverse characterizations Adaptive Cosine Estimator (MILMD-ACE), and Multiple Instance Learning for Multiple Diverse characterizations Spectral Match Filter algorithms (MILMD-SMF) \cite{zhong2019multiple}. This algorithm learns multiple target concepts by maximizing the collective dictionary's detection statistic across the positive bags while minimizing the detection across negative instances. A unique aspect of this algorithm is the assumption that each positive bag is constructed with multiple target types, which deviates from the traditional MIL framework.This approach is useful if two targets exist in every positive bag, but not in negative bags. 

The abovementioned algorithms have shown success at determining sub-pixel target signatures for target detection. However, every algorithm has limitations. Our proposed algorithms address four of these common limitations. First, the MTMI algorithms maximize the detection statistic of the dictionary with respect to expected target signatures provided from the training data. Second, the MTMI algorithms do not assume a set number of target signatures, but instead learns the appropriate number of target signatures. The abovementioned algorithms require that the number of targets, and often the number of background signatures, is known, which can be difficult without domain knowledge. Third, the uniqueness of returned target signatures is adjustable through a hyperparameter. The uniqueness hyperparameter allows for the multiple target signatures to be more similar or more distinct depending on the characteristics of the dataset. Lastly, the introduced algorithms are trained using a dot product in a transformed data space, which is fundamentally quick to compute, resulting in an efficient algorithm.

\subsection{Adaptive Cosine Estimator and Spectral Match Filter}

The Adaptive Cosine Estimator (ACE) and Spectral Match Filter (SMF) are detection statistics often used for hyperspectral sub-pixel target detection \cite{kraut1999cfar, kraut2001adaptive, basener2010clutter, nasrabadi2008regularized}. ACE performs detection by solely considering spectral shape, whereas SMF also considers magnitude. Assuming target signature $\mathbf{s}$ and unknown instance $\mathbf{x}$, the ACE and SMF detectors\cite{zare2018discriminative} can be written as  


\begin{equation}
    D_{ACE}(\mathbf{x}, \mathbf{s}) = \frac{\mathbf{s}^T\mathbf{P}^T}{||\mathbf{s}^T\mathbf{P}^T||} \frac{\mathbf{P}(\mathbf{x} - \boldsymbol{\mu_b})}{||\mathbf{P}(\mathbf{x} - \boldsymbol{\mu_b})||} 
    \label{ACESimplified}
\end{equation}

\begin{equation}
    D_{SMF}(\mathbf{x}, \mathbf{s}) = \frac{\mathbf{s}^T\mathbf{P}^T}{||\mathbf{s}^T\mathbf{P}^T||} \mathbf{P}(\mathbf{x} - \boldsymbol{\mu_b})
    \label{SMFSimplified}
\end{equation}

\begin{equation}
    \mathbf{P} = \mathbf{E}^{-\frac{1}{2}}\mathbf{U}^T
    \label{P}
\end{equation}

Here $D$ is the detection response of the given detector, either ACE or SMF. In equations \eqref{ACESimplified} and \eqref{SMFSimplified}, the detection statistics are shown as the dot product between the target signature and unknown sample in a whitened coordinate space. Here a translation is done using the mean of the background, $\boldsymbol{\mu_b}$, followed by a transformation using the eigenvectors, $\mathbf{U}$, and eigenvalues, $\mathbf{E}$, of the background covariance. When choosing which statistic is appropriate for a dataset, it is important to determine whether the spectral magnitude is necessary for a target’s discrimination from the background. When a target has a similar spectral shape to the background, but is brighter or darker (e.g., lower or higher reflectance), the SMF detection statistic would be appropriate.

\section{Methods}
\subsection{Multiple Target MIL}

Following the MIL framework, this algorithm assumes the data is grouped into bags with bag level labels \cite{dietterich1997solving}. With this, let $\mathbf{X} = \{\mathbf{x}_{1}, ..., \mathbf{x}_{N}\}$ be training data with each sample, $\mathbf{x}_{i}$ being a vector with dimensionality $D$. The data is grouped into $J$ bags $\mathbf{B} = \{\mathbf{B}_{1},..., \mathbf{B}_{J}\}$ with labels, $L = \{L_{1}, ..., L_{J}\}$, where $L_{j} \in \{0, 1\}$. A bag is considered positive, $\mathbf{B_{j}^{+}}$, with label $L_{j} = 1$, if there exists at least one instance, $\mathbf{x}_{i}$, in bag $j$ that is from the target class. Additionally, a bag is considered negative, $\mathbf{B_{j}^{-}}$, with label $L_{j} = 0$, if all instances in bag $j$ are from the background class. The number of instances in both positive and negative bags is variable.

In this work, the objective function of the original MI-ACE algorithm has been extended to include multiple target signatures. $K$ target signatures are in the dictionary, $\mathbf{S}$, shown here in Equation \eqref{MTMIACEDictionary}:
\begin{equation}
  \mathbf{S} = 
  \begin{bmatrix}
  \mathbf{s}_{1}, && \mathbf{s}_{2}, && ... && \mathbf{s}_{K}
  \end{bmatrix}
  \label{MTMIACEDictionary}
\end{equation}

The goal of MTMI-ACE and MTMI-SMF is to estimate the set of target signatures that maximize the detection statistic for the target instances in each positive bag and minimize the detection statistic over all negative instances. This is accomplished by maximizing the following objective function:

\begin{equation}
    M_+(\mathbf{B},\mathbf{S}) - M_-(\mathbf{B},\mathbf{S}) -  M_u(\mathbf{B},\mathbf{S}) \text{s.t.} \quad D(\mathbf{s}_{k}, \mathbf{s}_{k}) = 1
    \label{eq_MTMIObjFun}
\end{equation}

The objective function is comprised of three terms and a constraint \eqref{eq_MTMIObjFun}. The first term \eqref{M+} is the average detection statistic of the selected instances from each positive bag. This term also contains the target signatures to be learned. By including the max operation in the first term, each target signature will learn a particular target type. This operation is introduced because it is assumed every positive bag does not contain every target type. The dictionary of learned target signatures will maximize the objective function by individually maximizing a subset of positive bags. 

\begin{equation}
    M_+(\mathbf{B},\mathbf{S}) =  \underset{\textbf{S}}{\max} \quad \frac{1}{N^{+}} \sum_{j:L_{j} = 1} \underset{s_{k}\in \mathbf{S}}\max (D(\mathbf{x}_{j,k}^{*}, \mathbf{s}_{k}))  
    \label{M+}
\end{equation}
\noindent
Here $N^{+}$ is the number of positive bags and $\mathbf{x}_{j,k}^{*}$ is the selected instance from the positive bag $\mathbf{B_{j}^{+}}$ that is most likely a target instance in the bag. The selected instance $\mathbf{x}_{j,k}^{*}$ is identified as the point in bag $j$ with the maximum detection statistic given a target signature, $\mathbf{s}_{k}$. 

\begin{equation}
  \mathbf{x}_{j,k}^{*} = \underset{x_{j}\in \mathbf{\mathbf{B}_{j}^{+}}}\argmax D(\mathbf{x}_{j}, \mathbf{s}_{k})
  \label{eq_xstar}
\end{equation}

The second term \eqref{M-} is the average detection statistic of the negative bag's instances and the target signatures to be learned. This term discourages learning any target signature that is similar to the background. By including the outer most sum, each negative bag has an equal weight in the objective function. Here $N^{-}$ is the number of negative bags respectively, and $N_{j}^{-}$ is the number of instances in negative bag $j$.

\begin{equation}
    M_-(\mathbf{B},\mathbf{S}) =  \frac{1}{K} \sum_{k=1}^{K} \frac{1}{N^{-}} \sum_{j:L_{j} = 0} \frac{1}{N_{j}^{-}} \sum_{x_{i} \in B_{j}^{-}} D(\mathbf{x}_{i}, \mathbf{s}_{k}) 
    \label{M-}
\end{equation}

The third term, known as the uniqueness term \eqref{Mu}, is introduced to encourage the algorithm to learn distinct target signatures by penalizing the algorithm for similar target signatures. The uniqueness term has a hyperparameter weight, $\alpha$, which changes how similar the target signatures can be. The larger the weight, the more the algorithm will be encouraged to learn different signatures. 

\begin{equation}
    M_u(\mathbf{B},\mathbf{S}) =  - \frac{\alpha}{{K \choose 2}}\sum_{k=1}^{K-1} \sum_{l=k+1}^{K} D(\mathbf{s}_{k}, \mathbf{s}_{l}) \quad
    \label{Mu}
\end{equation}

Finally, the constraint, $D(\mathbf{s}_{k}, \mathbf{s}_{k}) = 1$, is included to restrict the algorithm from learning target signatures of erroneously large magnitude. 
  
The MTMI-ACE and MTMI-SMF algorithms have two primary steps, initialization and optimization. The initialization process uses a greedy approach along with clustering to aid in computation complexity and representative target diversification. The optimization process learns the number of needed signatures to optimally describe the target class while generalizing the signatures considering all of the positive bags. These steps are described in the following sections. The algorithm pseudocode is provided in Algorithm \ref{MIMIACEAlg}.

\begin{algorithm}
\caption{MTMI-ACE and MTMI-SMF}\label{MIMIACEAlg}
\begin{algorithmic}[1]
\State Compute $\boldsymbol{\mu_{b}}$, $\mathbf{E}$, and $\mathbf{U}$
\State Subtract $\boldsymbol{\mu_{b}}$ and whiten $\mathbf{X}$, Equation \eqref{P}
\If {ACE} 
\State Normalize $\mathbf{X}$
\EndIf
\State K-Means cluster instances in positive bags
\State Greedily initialize $\mathbf{S}$, as the cluster centers that maximize the objective function, Equation \eqref{eq_MTMIObjFun}
\Repeat
  \State Update the set of $\mathbf{x}_{j,k}^{*}$ for each $\mathbf{s}_{k}$, Equation \eqref{eq_xstar}
  \State Determine indicator $\delta_{j,k}^{+}$ for each $\hat{\hat{\mathbf{s}}}_{k}$, Equation \eqref{MTMIACEOptimizationBagIdentifier}
  \For{$k = 1$ to $K$}
    \If{sum($\delta_{k}^{+}$) = 0}
      \State Remove $\mathbf{s}_{k}$ from $\mathbf{S}$
    \EndIf
    \State Update $\hat{\hat{\mathbf{s}}}_{k}$, Equation \eqref{MTMIACEUpdate1}
  \EndFor
\Until Stopping criterion reached
\State Normalize and de-whiten $\hat{\hat{\mathbf{s}}}_{k}$, Equation \eqref{dewhiteeq}
\State \Return all optimized $\mathbf{s}_{k}$ as $\mathbf{S}$
\end{algorithmic}
\label{Alg:MTMIACE}
\end{algorithm}

\subsection{Target Dictionary Initialization}
To reduce the computation complexity of the initialization process, the K-Means clustering algorithm \cite{macqueen1967some} is used to aid in target concept initialization. K-Means is used by clustering all of the data, regardless of bag structure, into $C$ clusters. Then, the cluster centers that maximize the objective function in Equation \eqref{eq_MTMIObjFun} are iteratively selected until $K$ signatures have been added to the target dictionary. In this process, the algorithm is greedily selecting the next best target concept for initialization. As long as the number of clusters, $C$, and the number of iterations $i$, remains small, the K-Means approach will have a lower computational cost than searching through all of the positive instances. Using K-Means, the algorithm only needs to search through $C$ candidates instead of $N^{+}$ candidates to initialize a target signature.

\subsection{Target Dictionary Optimization}
The update equation to perform target concept optimization is calculated for each target concept individually. Each target concept is updated iteratively, along with the other target concepts. This solution is closed-form. The optimization will produce the same result given a fixed set of positive bags, negative bags, and initialized target signatures \cite{zare2018discriminative}. This assumes no cycles occur during optimization, and the max number of iterations is not reached. The update equation to perform target concept optimization is derived by maximizing the objective function for the target signatures, $\hat{\hat{\mathbf{s}}}_{k}$. The derivation of the update equation is included in Appendix \ref{sec:appendixA}. The resulting update equation is:
\begin{equation}
    \hat{\hat{\mathbf{s}}}_{k} = \frac{\hat{\hat{\mathbf{t}}}}{\left \| \hat{\hat{\mathbf{t}}} \right \|} 
    \label{MTMIACEUpdate1}
\end{equation}
where 
\begin{eqnarray}
    \hat{\hat{\mathbf{t}}} &=& \frac{1}{N_{k}^{+}} \sum_{j:L_{j} = 1} \delta_{j,k}^{+} \hat{\hat{\mathbf{x}}}^{*}_{j,k} \\ &-& \frac{1}{N^{-}} \sum_{j:L_{j} = 0} \frac{1}{N_{j}^{-}} \sum_{\hat{\hat{x}}_{i} \in B_{j}^{-}} \hat{\hat{\mathbf{x}}}_{i} - \frac{\alpha}{(K-1)} \sum_{l, l \neq k} \hat{\hat{\mathbf{s}}}_{l}
\end{eqnarray}
    
\begin{equation}
    \hat{\hat{\mathbf{x}}} = \frac{\mathbf{P}(\mathbf{x} - \boldsymbol{\mu_b})}{||\mathbf{P}(\mathbf{x} - \boldsymbol{\mu_b})||} 
\end{equation}

where
\begin{equation}
  \delta_{j,k}^{+} =
  \begin{cases}
    1 & \text{if} \: z > 0 \\
    0 & \text{otherwise}
  \end{cases}
  \label{MTMIACEOptimizationBagIdentifier}
\end{equation}

\begin{equation}
    z = D\big(\mathbf{x}^{*}_{j,k}, \mathbf{s}_{k}\big) - D\big(\mathbf{x}^{*}_{j,l}, \mathbf{s}_{l}\big)
\end{equation}


This update equation is interpretable. The first term is the average of the selected positive instances assigned to the same target type as $\mathbf{\hat{\hat{s}}}_{k}$. The selected positive instances are determined by computing the positive bag identifiers $\delta^{+}$ for each of the target signatures. The second term is the average of all negative instances, and the third term is the average of all other target signatures. This result is also intuitive. A learned target signature will be an average of the target type being optimized, dissimilar from the background, and pushed away from the other target signatures to encourage target signature uniqueness. Depending on the application, $\alpha$ is configurable to allow for more distinct or similar target signatures. If $\alpha$ is increased, more distinct target signatures are allowed. Finally, all target signatures are optimized simultaneously until the bag identifiers, $\delta^{+}$, and each target signature's bag representatives, $\hat{\hat{\mathbf{x}}}^{*}_{k}$, remain the same across subsequent iterations, or the max number of iterations is reached. At this point, all of the positive instances pertaining to each target concept have been determined, and the algorithm has optimized the signatures. The final step is to normalize and de-whiten target signatures using:
\begin{equation}
    \mathbf{s}_{k} = \frac{\mathbf{t}_{k}}{\vert\vert \mathbf{t}_{k} \vert\vert}, \mathbf{t}_{k} = \mathbf{P}^{-1} \hat{\hat{\mathbf{s}}}_{k}
    \label{dewhiteeq}
\end{equation}

\subsection{Learning Number of Target Concepts}
During optimization, the number of target concepts is estimated iteratively by removing unnecessary target signatures. Target signatures are removed by observing the bag identifiers, in \eqref{MTMIACEOptimizationBagIdentifier}, for each of the $k$ target signatures during the iterations of optimization. If the bag identifiers for the $k^{th}$ target are $0$ for all $j$ positive bags, then the $k^{th}$ target signature will be dropped from the set of target signatures $\mathbf{S}$. Namely, the $k^{th}$ signature will be dropped when the detection similarity between the $k^{th}$ target signature and all $j$ positive bag representatives, $\hat{\hat{\mathbf{x}}}^{*}_{k}$, are smaller than all other target signatures' detection similarities to their corresponding $j$ bag representatives. With this, MTMI-ACE and MTMI-SMF remove the need for domain-specific knowledge for how many target signatures may exist; while still being adjustable by changing the value of $\alpha$ to encourage more or less target signature uniqueness. Where a lower $\alpha$ value can potentially lead to more target signatures to be estimated since signatures are allowed to be more similar.  

\section{Experiments}
In the following, MTMI-ACE and MTMI-SMF are evaluated and compared to several MIL framework methods using simulated data and two real hyperspectral datasets. The simulated data experiments are included to illustrate the properties of MTMI-ACE and MTMI-SMF, providing insight into how and when the methods are effective. The hyperspectral datasets are included to illustrate how MTMI-ACE and MTMI-SMF perform in real-world scenarios with two different bagging methods (Figure \ref{fig_bagexample}). In this section, we will compare our proposed algorithms with other MIL algorithms from the literature including: Multiple Instance Learning for Multiple Diverse characterizations (MILMD) \cite{zhong2019multiple}, Multiple Instance Adaptive Cosine Estimator (MI-ACE) \cite{zare2018discriminative}, Multiple Instance Spectral Matched Filter (MI-SMF) \cite{zare2018discriminative}, Multiple Instance Hybrid Estimator (MI-HE) \cite{jiao2018multiple}, and extended Functions of Multiple Instances (eFUMI) \cite{jiao2015functions}. These algorithms were selected due to their relevance, prevalence in the literature, or their recent development. 

\subsection{Simulated Data: Single Target}
\subsubsection{Experimental Dataset}
Simulated data were generated from four spectra selected from the ECOSTRESS Spectral Library \cite{Meerdink2019b}, formally known as the ASTER Spectral Library \cite{Baldridge2009}. Those five spectra were from the rock class (basalt, pyroxenite, verde antique, phyllite, and slate) and had 211 bands ranging from 400 - 2500nm. The simulated dataset was generated following steps and code detailed in \cite{Bockinsky2019, jiao2015functions}. In this experiment, the simulated dataset was created using two target signatures: basalt and verde antique. All other spectra (three classes) were used as background. The parameters used to develop the simulated dataset were 10 positive bags, 20 negative bags, 500 points in each bag, 250 target points in each positive bag, 0.3 mean target proportion, and 20 signal to noise ratio. Two simulated datasets were generated using those parameters, with one designated for training and the other for testing. This experimental design was repeated for ten iterations. For these simulated experiments, algorithms were evaluated on this data using the Normalized Area Under the receiver operating characteristic Curve (NAUC) in which the area was normalized out to a false alarm rate (FAR) of 1x$10^{-3}$ false alarms/m$^{2}$.

\begin{figure*}[h!]
    \centering
    \includegraphics[width=\textwidth]{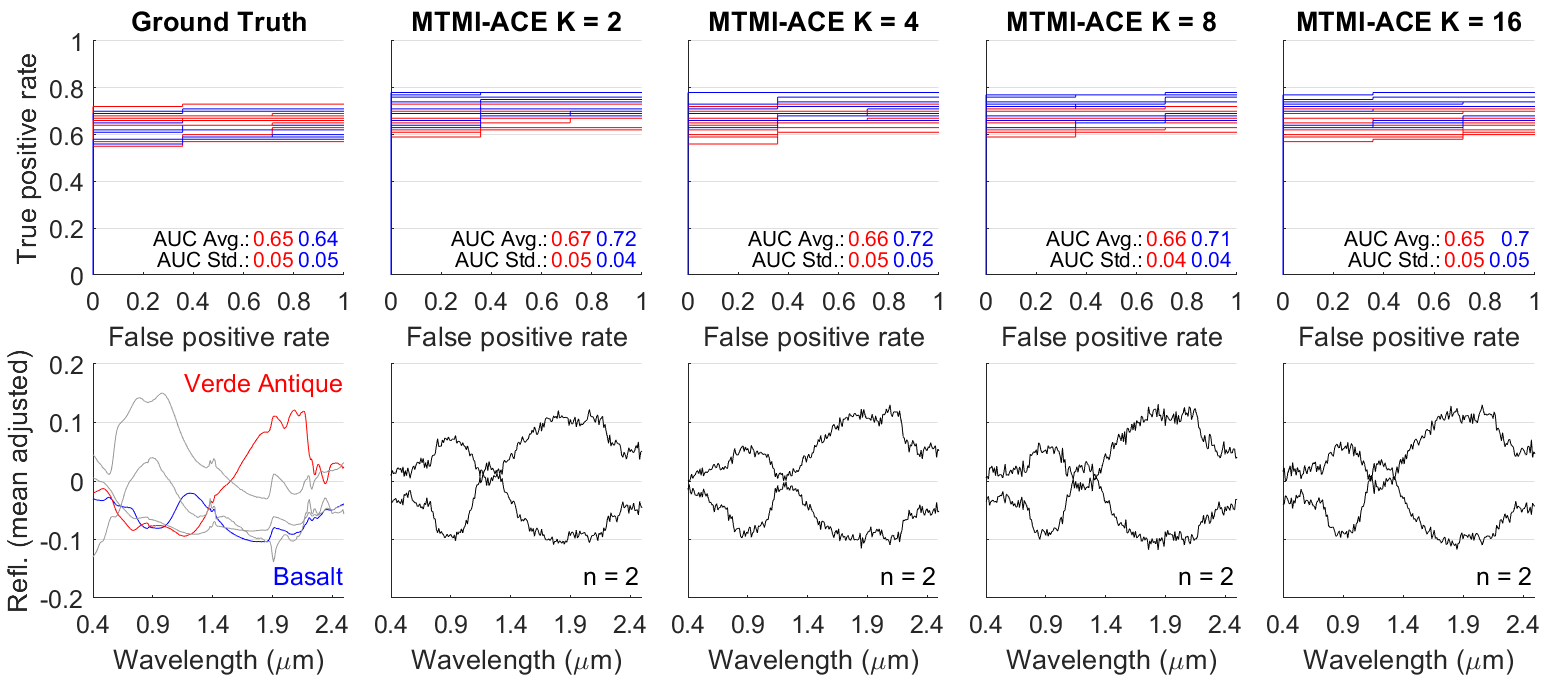}
    \caption{MTMI-ACE ROC curves (top) across the ten dataset simulations and changing \textit{K} values and target signatures from the highest performing iteration (bottom). The first column shows the target detection results using the original target signatures and original signatures, after whitening, with the two targets in red and blue with background signatures in gray. The false-positive rate scale is shown as 1x10$^{-3}$.}
    \label{fig_SimK}
\end{figure*}

\begin{figure*}[h!]
    \centering
    \includegraphics[width=\textwidth]{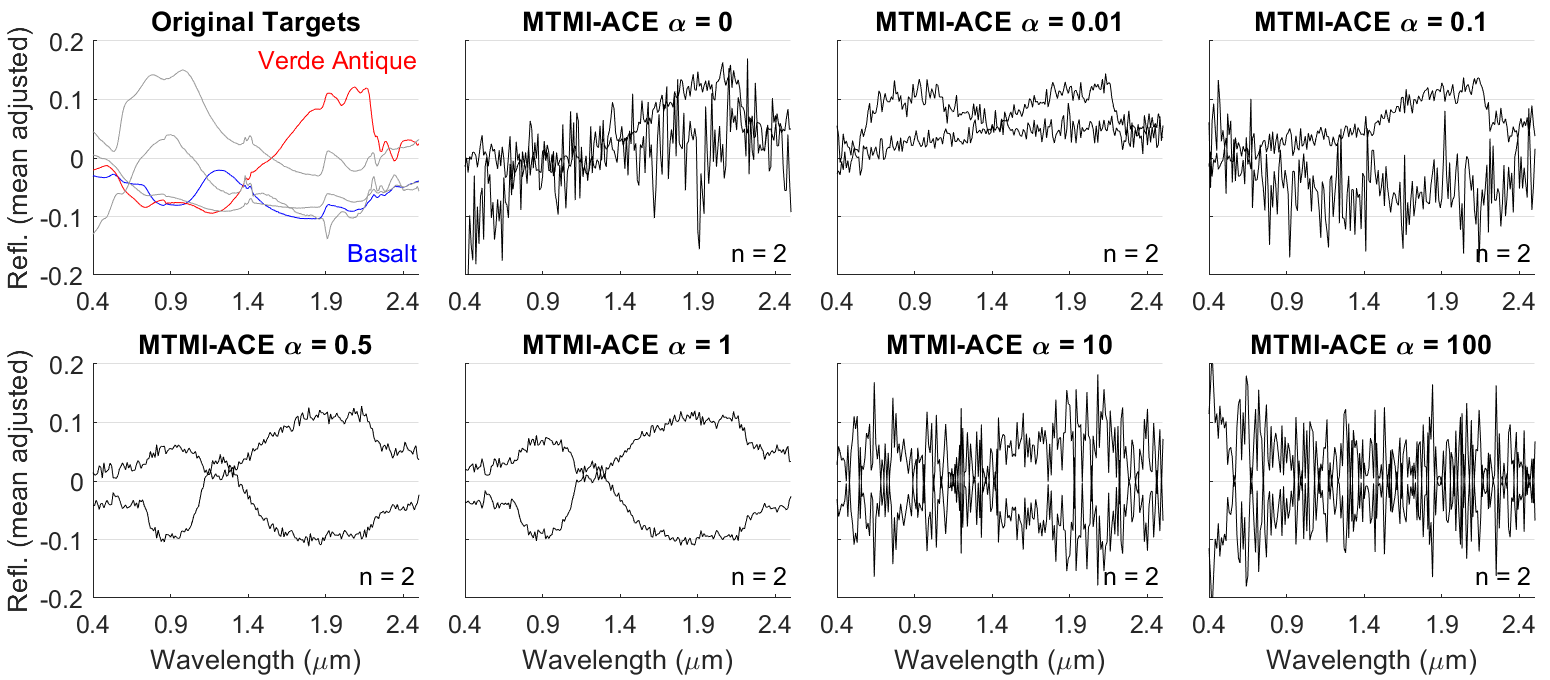}
    \caption{The MTMI-ACE target signatures from the highest performing iteration across different $\alpha$ values. The first panel shows the original signatures, after whitening, with the two targets in red and blue with background signatures in gray. }
    \label{fig_SimAlpha}
\end{figure*}

\subsubsection{Effects of K}
The \textit{K} parameter in the Multi-Target Multiple Instance algorithm controls the number of initial target signatures. Too few targets may not capture the spectral variability in the target’s pixels. A large \textit{K} parameter slows down computation. Often the user does not know what the appropriate \textit{K} value is for their dataset. In this experiment, all MTMI-ACE and MTMI-SMF parameters were kept constant ($\alpha$ = 0.5) while the \textit{K} value was changed from 2, 4, 8, and 16. 

With changing \textit{K}, the developed target signatures are similar, and the total returned number of targets was consistent (Figure \ref{fig_SimK}). Some iterations did show a larger number of returned targets with \textit{K} equal to 8 and 16, but for the most iterations, the total returned number of targets was 2 indicating the stability in the ability to estimate the number of target signatures. The target detection performance shown in the ROC curves and NAUC values do not show much variability with changing \textit{K}.

Another benefit of the MTMI algorithms is the fact that the returned target signatures are interpretable. The returned signatures are similar to the original signatures, but with a few key differences (Figure \ref{fig_SimK}). The MTMI algorithms maximize differences between the targets and the background, so spectral features that are found beneficial for separation are exaggerated. For example, the peak in the verde antique spectra between 1.5 - 2.3 $\mu$m is a unique feature of this class. The returned MTMI target signature reflects this difference showing elevated reflectance values in this spectral range. These returned target signatures could be used to link back to specific biochemistry or physical properties of the target class. It is often useful in remote sensing applications to know what wavelengths make a target class different from background classes.  
  
 \begin{figure*}
    \centering
    \includegraphics[width=\textwidth]{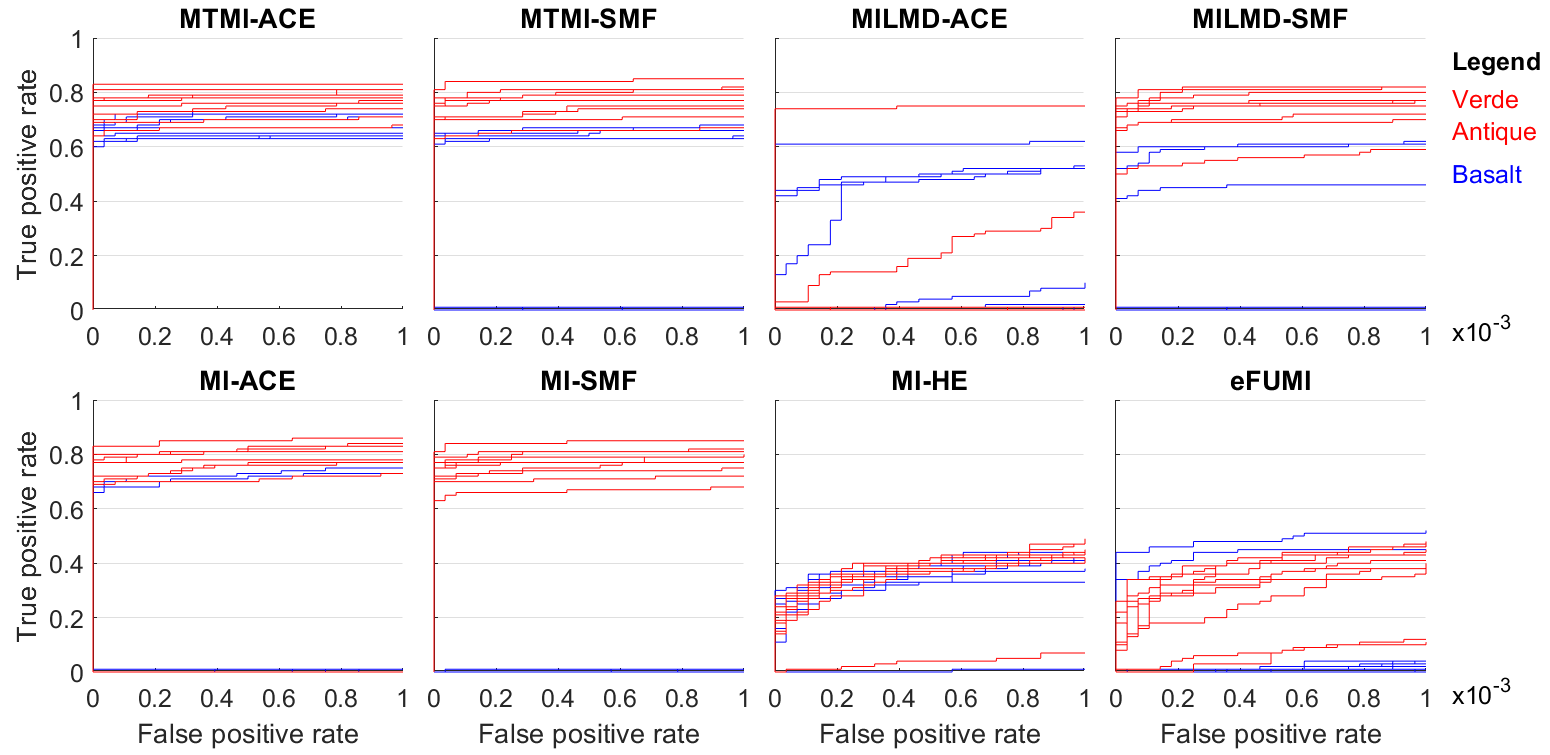}
    \caption{The two target simulated dataset ROC curves for each of the ten dataset simulations across the nine target detection algorithms. The false-positive rate scale is shown as 1x10$^{-3}$. At this range, some of the iterations had a zero true-positive rate.}
    \label{fig_SimTargets}
\end{figure*}

\subsubsection{Effects of $\alpha$}
The $\alpha$ parameter controls the similarity or diversity of target signatures obtained from MTMI-ACE and MTMI-SMF. A smaller $\alpha$ will allow for more similar target signatures, while a larger $\alpha$ will force target signatures to be more diverse. Changing the $\alpha$ parameter can have a large effect on returned target signatures depending on the spectral variability in the dataset. In this experiment, all MTMI-ACE and MTMI-SMF parameters are kept constant (\textit{K} = 2) while the $\alpha$ value is changed from 0, 0.01, 0.1, 0.5, 1, 10, to 100. 
	
For this dataset, the best $\alpha$ for returning interpretable target signatures was 0.5 and 1 (Figure \ref{fig_SimK}). After a value of 1, the signatures become noisy and do not reflect any features from the original signatures. As mentioned above, the $\alpha$ parameter controls the amount of diversity between developed signatures. An $\alpha$ of 0 does not use the third term of the objective function at all and does not encourage diversity among the multiple targets. As $\alpha$ increases, the target signatures become increasingly different from each other until the $\alpha$ increases too much. However, the target detection performance shown in the ROC curves and NAUC values do not show much variability with changing $\alpha$. It is essential to examine the target signatures to ensure they are making physical sense and retain confidence in detection results. 

When selecting an $\alpha$, choose values less than 1 and greater than 0. The appropriate $\alpha$ is dependent on the spectral differences between target and background classes. For targets that are similar spectrally or exhibit significant overlap between targets, a smaller $\alpha$ would ensure that developed target signatures would reflect patterns in the dataset. While not seen in this dataset, a larger $\alpha$ can reduce the number of developed target signatures so that diversity is maximized. If multiple targets are needed for accurate detection, a smaller $\alpha$ will ensure multiple target signatures will be developed because the signatures are allowed to be more similar. 

\begin{table}
    \setlength\extrarowheight{2pt}
    \centering
    \begin{tabular}{|c c c|}
        \hline
         Algorithm & Basalt & Verde Antique\\
         \hline
         MTMI-ACE & \textbf{0.652 (0.019)} & \textbf{0.784 (0.056)}\\
         MTMI-SMF & \underline{0.318 (0.333)} & \underline{0.741 (0.054}\\
         MILMD-ACE &  0.164 (0.237) & 0.078 (0.233)\\
         MILMD-SMF & 0.155 (0.250) & 0.712 (0.077)\\
         MI-ACE  & 0.138 (0.287) & 0.608 (0.323)\\
         MI-SMF  & 0.001 (0.002) & 0.745 (0.052)\\
         MI-HE   & 0.223 (0.087) & 0.211 (0.077)\\
         eFUMI   & 0.080 (0.167) & 0.167 (0.105)\\
         \hline
    \end{tabular}
    \caption{Normalized average area under curve (NAUC) results, with standard deviation in parentheses, for the simulated dataset with two targets (Basalt and Verde Antique). Best results shown in bold, second-best results are underlined.}
    \label{tab:SIM2results}
\end{table}

\subsubsection{Target Detection Results}
Using the two target simulated dataset, we compared MTMI with six other multiple instance target detection algorithms in the literature. For this experiment, MTMI-ACE and MTMI-SMF parameters were fixed with $\alpha$ = 1 and \textit{K} = 4. Although we know this dataset only has two targets, four potential targets were allowed to demonstrate that MTMI algorithms return the appropriate number of targets for the dataset. This flexibility eases the restrictions on the user-defined \textit{K} parameter. Other algorithm parameters are found in Appendix B. 

Figure \ref{fig_SimTargets} shows the ROC curves for the two classes across algorithms, and Table \ref{tab:SIM2results} shows the average NAUC and standard deviations from the ten dataset simulations. The results from this simulation become more evident when compared to the original target signatures, shown in panel 1 of Figure \ref{fig_SimK}. Compared to the other spectra, verde antique has a unique spectrum specifically in the 1.5 - 2.3 $\mu$m spectral range, while basalt shares more similar features. The more distinct the target spectrum, the easier it is for the target detection algorithms to determine the appropriate target signature. All the algorithms were able to determine an appropriate signature for verde antique, resulting in higher accuracy for that target class. However, the main deviation among algorithm performance happened when the algorithms tried to determine the appropriate signature for basalt and were not able to capture the spectral signature. 

MTMI-ACE performs well for detecting basalt and verde antique compared to many of the other algorithms. Notably, the NAUC values are good for both target types compared to MI-ACE were one target performing significantly better. Looking at the NAUC, it becomes clear that the ability to determine multiple targets leads to an increase in detection compared to MI-ACE and MI-SMF, which can only return one target signature. These algorithms were able to detect the verde antique target with high accuracy, but not the basalt target because only a verde antique target was developed. MI-ACE returned target signatures similar to verde antique because it was distinct from the other targets, while basalt was too similar. Another recently released multiple target MIL algorithm, MILMD-SMF, also detected the two targets well but returned target signatures that favored the verde antique target leading to a decrease mean NAUC for basalt target detection. MILMD-ACE shows large variability across iterations. These algorithms were developed with the assumption that both targets exist in each positive bag instead of the traditional approach of a single target, which is how this dataset was created. The MI-HE algorithm showed high consistency across random iterations, while many other algorithms showed high variability in results and also learned target signatures.

\begin{figure}
    \centering
    \includegraphics[width=3.4in]{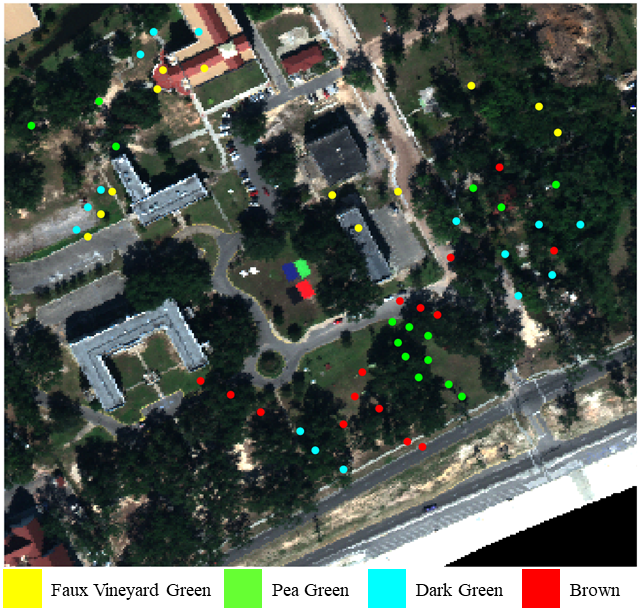}
    \caption{MUUFL Gulfport dataset Red/Green/Blue image and the 57 target locations.}
    \label{fig_MUUFLStudySite}
\end{figure}

\subsection{MUUFL Gulfport Data}
\subsubsection{Experimental dataset}
The MUUFL Gulfport Hyperspectral dataset was used to perform experiments on hyperspectral data that contained sub-pixel targets (Figure \ref{fig_bagexample}). This dataset was collected over the University of Southern Mississippi-Gulfpark Campus with 1m spatial resolution and 72 bands ranging between 367.7nm to 1043.4nm. In this study, we are using two images (flight 1 and flight 3) which cover the same spatial area but were flown approximately 10 minutes apart. These images contain 57 human-made targets made of cloth panels in four different colors: brown (15 panels), dark green (15 panels), faux vineyard green (12 panels), and pea-green (15 panels). The targets’ spatial location are shown as scattered points over a Red/Green/Blue image of the scene in Figure \ref{fig_MUUFLStudySite}. This dataset is a very challenging target detection task as trees partially or fully occlude many of the targets. Furthermore, the targets vary in size and could be 0.25m\textsuperscript{2}, 1m\textsuperscript{2} and 9m\textsuperscript{2}. Thus, a target that has 0.25m\textsuperscript{2} covers at most a 0.25 proportion of the pixel signature if the pixel falls directly on the target. However, many of these targets straddle multiple pixels and are occluded, resulting in a highly mixed, sub-pixel target detection task. A bag included all pixels in a 5 x 5 rectangular region around each ground truth point. The GPS device used to record the ground truth locations had a maximum of 5m accuracy, which would result in a 10 x 10 rectangular region. However, the GPS accuracy on the day of collection was around 2-3m, which is why the size of 5 x 5 was chosen. The remaining area that did not contain the target class was grouped into one big negative bag. Two iterations were run in which flight 1 was selected for training and flight 3 for testing, and vice versa. The target types were iterated through, and in each iteration, a single target type was selected as a positive bag, and all other image pixels were selected as a negative bag. Thus, there are 57 positive bags in each training set in this experiment. For this experiment, algorithms were evaluated on this data using the Normalized Area Under the receiver operating characteristic Curve (NAUC) in which the area was normalized out to a false alarm rate (FAR) of 1x$10^{-3}$ false alarms/m$^{2}$ \cite{Glenn2016a}. 

\begin{figure}
    \centering
    \includegraphics[width=3.4in]{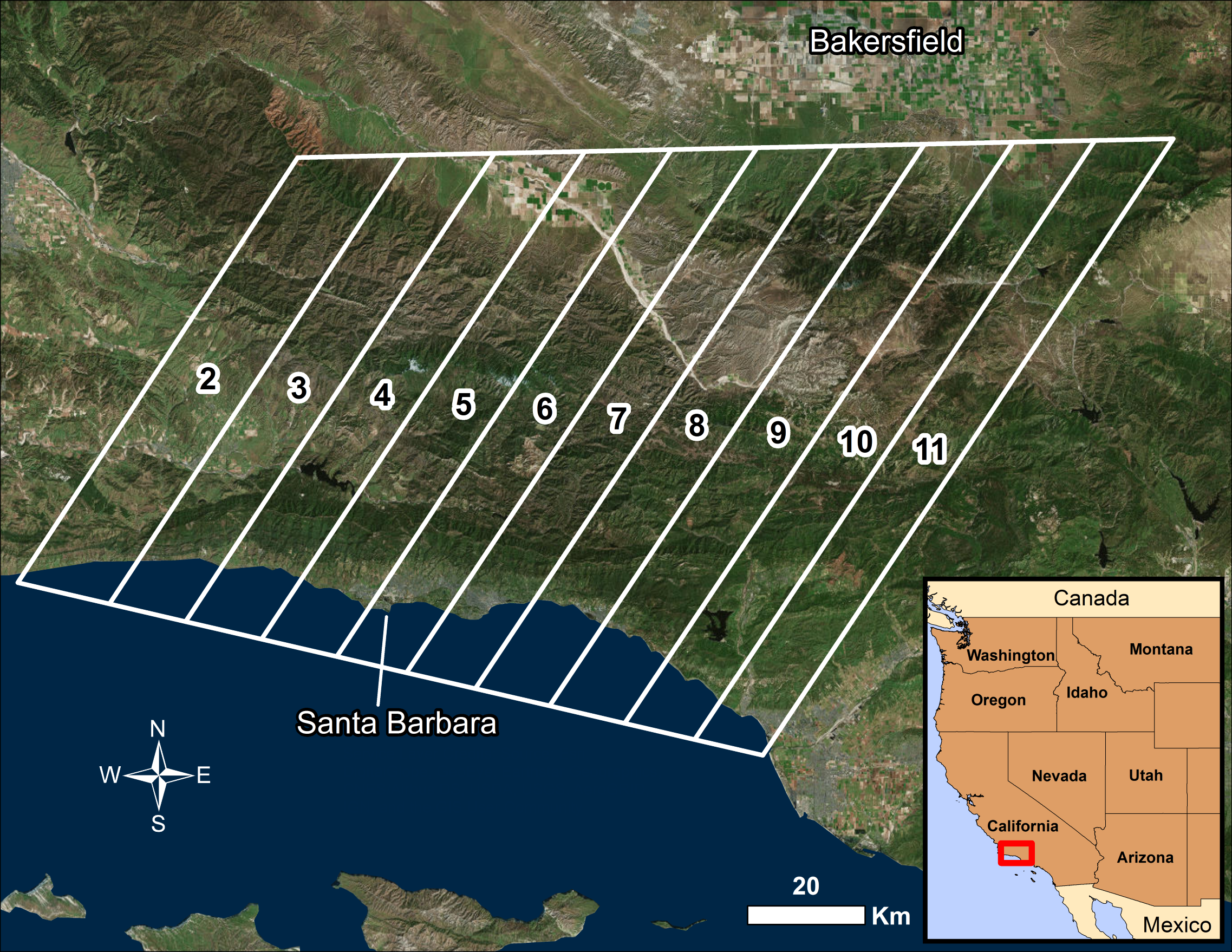}
    \caption{Santa Barbara flight box and the HyspIRI Airborne Preparatory campaign flight lines used \cite{Meerdink2019b}.}
    \label{fig_SKMStudySite}
\end{figure}

\subsubsection{Target Detection Results}
\begin{figure*}[!h]
    \centering
    \includegraphics[width=\textwidth]{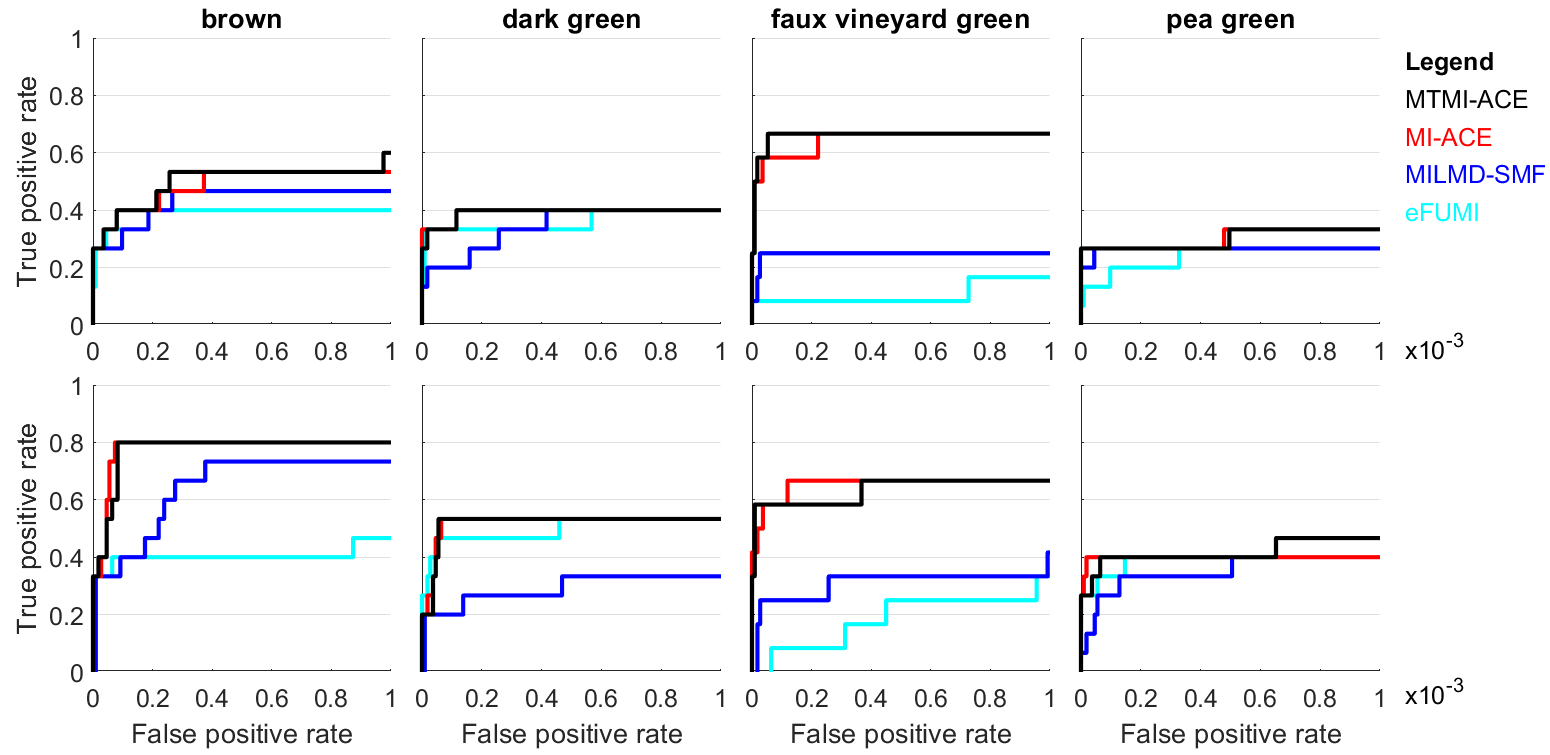}
    \caption{The MUUFL Gulfport dataset ROC curves for a subset of algorithms. The top row shows results from training with flight line 1 and testing on flight line 3. The bottom row shows results from training with flight line 3 and testing on flight line 1. The false-positive rate scale is shown as 1x10$^{-3}$. }
    \label{fig_MUUFLresults}
\end{figure*}

As mentioned above, the MUUFL Gulfport dataset represents a challenging sub-pixel detection environment due to many trees covering targets. Additionally, these targets are spectrally similar to the background because they are cotton fabric (which exhibit vegetation spectral features) and are similar colors (e.g., shades of green and brown). All these dataset characteristics result in lower detection results compared to other datasets. However, it does provide the opportunity to test the capabilities of the detection algorithms in a challenging scenario.  The MUUFL Gulfport dataset NAUC results are shown in Table \ref{tab:MUUFLresults} for the two training and testing splits across flight lines and Figure \ref{fig_MUUFLresults} shows the four target ROC curves for a subset of algorithms. 

MTMI-ACE ranked among the highest performing across the training/testing split. In cases were MTMI-ACE/MTMI-SMF were not the highest performing, MI-ACE algorithm generally performed the best, demonstrating that for this dataset, a single target was sufficient for accurate detection. In many cases, multiple target signatures did not yield higher detection rates, as is evident by MI-ACE and MI-SMF performing equally well or better for classes. The MILMD-SMF algorithm, which is most similar to our proposed algorithm, did have comparable results to the MTMI algorithm. However, the MILMD-ACE algorithm was not able to determine an appropriate target signature for many of the classes. MI-HE and eFUMI algorithms also performed comparably, but slightly decreased accuracies. However, these algorithms took much longer to computationally execute and have more user-defined parameters that can impact the final results.  

\begin{table*}
    \centering
    \setlength\extrarowheight{2pt}
    \begin{tabular}{|c c c c c c c c c c|}
        \hline
          & \multicolumn{4}{c}{Train on flight 1; Test on flight 3} & & \multicolumn{4}{c|}{Train on flight 3; Test on flight 1} \\
          \cline{2-5}
          \cline{7-10}
         & Brown & Dark Green & Faux V. Green & Pea Green & & Brown & Dark Green & Faux V. Green & Pea Green \\
         \hline
         MTMI-ACE & \textbf{0.496}&	\underline{0.391} &	\textbf{0.659}&	\underline{0.300}&&	\underline{0.772} &	\underline{0.518} &	\underline{0.634} &	\textbf{0.416} \\
         MTMI-SMF & 0.452& 0.365&	0.472&	0.267&&	0.667&	0.417&	0.492&	0.396 \\
         MILMD-ACE & 0.391&	0.013&	0.000&	0.000&&	0.742&	0.000&	0.000&	0.267\\
         MILMD-SMF & 0.430&	0.343&	0.246&	0.264&&	0.638&	0.291&	0.307&	0.350\\
         MI-ACE & \underline{0.486} &	\textbf{0.392}&	\underline{0.643} &	\textbf{0.301}&&	\textbf{0.777}&	\textbf{0.519}&	\textbf{0.652}&	\underline{0.398}\\
         MI-SMF & 0.452&	0.364&	0.468&	0.267&&	0.668&	0.406&	0.510&	0.396\\
         MI-HE & 0.433 & 0.379 & 0.104 &0.267 && 0.710 & 0.360 & 0.111 & 0.266\\
         eFUMI & 0.383&	0.360&	0.106&	0.238&&	0.404&	0.497&	0.185&	0.387\\
         \hline
    \end{tabular}
    \caption{Normalized Area Under the receiver operating characteristic Curve (NAUC) in which the area was normalized out to a false alarm rate of 1x10$^{-3}$ false alarms/m$^{2}$ for the MUUFL Gulfport dataset. Best results shown in bold, second-best results are underlined.}
    \label{tab:MUUFLresults}
\end{table*}

\subsection{AVIRIS Santa Barbara Data}
\subsubsection{Experimental dataset}
The Airborne Visible/Infrared Imaging Spectrometer (AVIRIS) Santa Barbara dataset was used to perform experiments on real hyperspectral data that contained training data in the form of polygons that may not contain all pure pixels (Figure \ref{fig_bagexample}). The imagery was collected with the AVIRIS sensor as part of the HyspIRI Airborne Preparatory Campaign on April 16, 2014 \cite{Lee2015}. AVIRIS measures 224 bands of radiance between 360 and 2500 nm with a full width at half-maximum of 10 nm \cite{Green1998}. This study uses a spatial subset of imagery from the Santa Barbara flight box, which includes ten of the eleven flight lines that were acquired with a 35º northeast-southwest orientation and 18 m spatial resolution (Figure \ref{fig_SKMStudySite}). These ten flight lines cover a diverse landscape that is approximately 12,980 km\textsuperscript{2}. For more information about imagery pre-processing and development of the training dataset, please refer to \cite{Meerdink2019b}. The original training data was collected to classify plant species, but this study grouped plant species into their plant functional types (PFTs). This resulted in nine classes of PFTs: annual herb (AH), deciduous broadleaf tree (DBT), deciduous shrub (DS), evergreen broadleaf tree (EBT), evergreen broadleaf shrub (EBS), evergreen needleleaf shrub (ENS), evergreen needleleaf tree (ENT), rock/soil (RS), and urban (URB). The training dataset comprised of spatial polygons designating where on the landscape `pure' patches of species existed. These locations were identified in the field and using AVIRIS and National Agriculture Imagery Program (NAIP) imagery. However, it is often challenging to find 100\% pure patches of species on the landscape, so patches having greater than 75\% single species composition were recorded. 

Each polygon was treated as a bag, and often, the pixels in a positive bag do not all belong to the target class. Often in traditional classifiers, these non-target pixels will add too much variability and confusion. Data were split into training and testing using 5-Fold cross-validation. Iterating through each class, all polygons matching that class label were selected as positive bags, while all other polygons were chosen as negative bags. For MTMI-ACE/MTMI-SMF parameters, the background mean and covariance were calculated from all pixels in the reference library, the \textit{K} was 15, and $\alpha$ was 1. For this experiment, algorithms were evaluated on this data using the Normalized Area Under the receiver operating characteristic Curve (NAUC) in which the area was normalized out to a false alarm rate (FAR) of 1x$10^{-2}$ false alarms/m$^{2}$.

\begin{table*}
    \includegraphics[width=\linewidth]{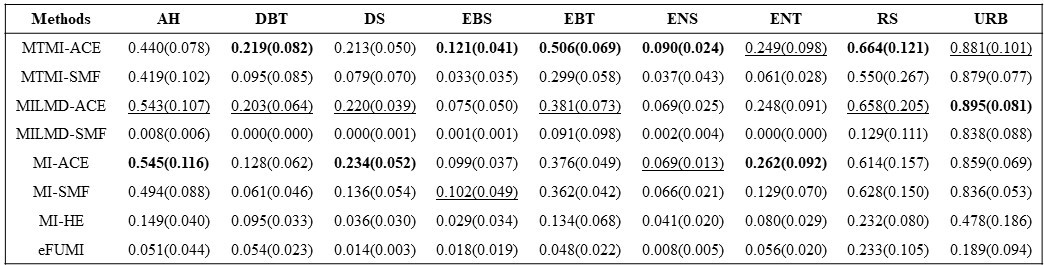}
    \caption{Averaged normalized area under the curve (NAUC) with standard deviation in parentheses for each of the nine classes in the AVIRIS Santa Barbara dataset across the different methods. Best results (based on average NAUC) are shown in bold; second-best results are underlined.}
    \label{tab:AVIRISresults}
\end{table*}

\begin{figure*}
    \centering
    \includegraphics[width=\textwidth]{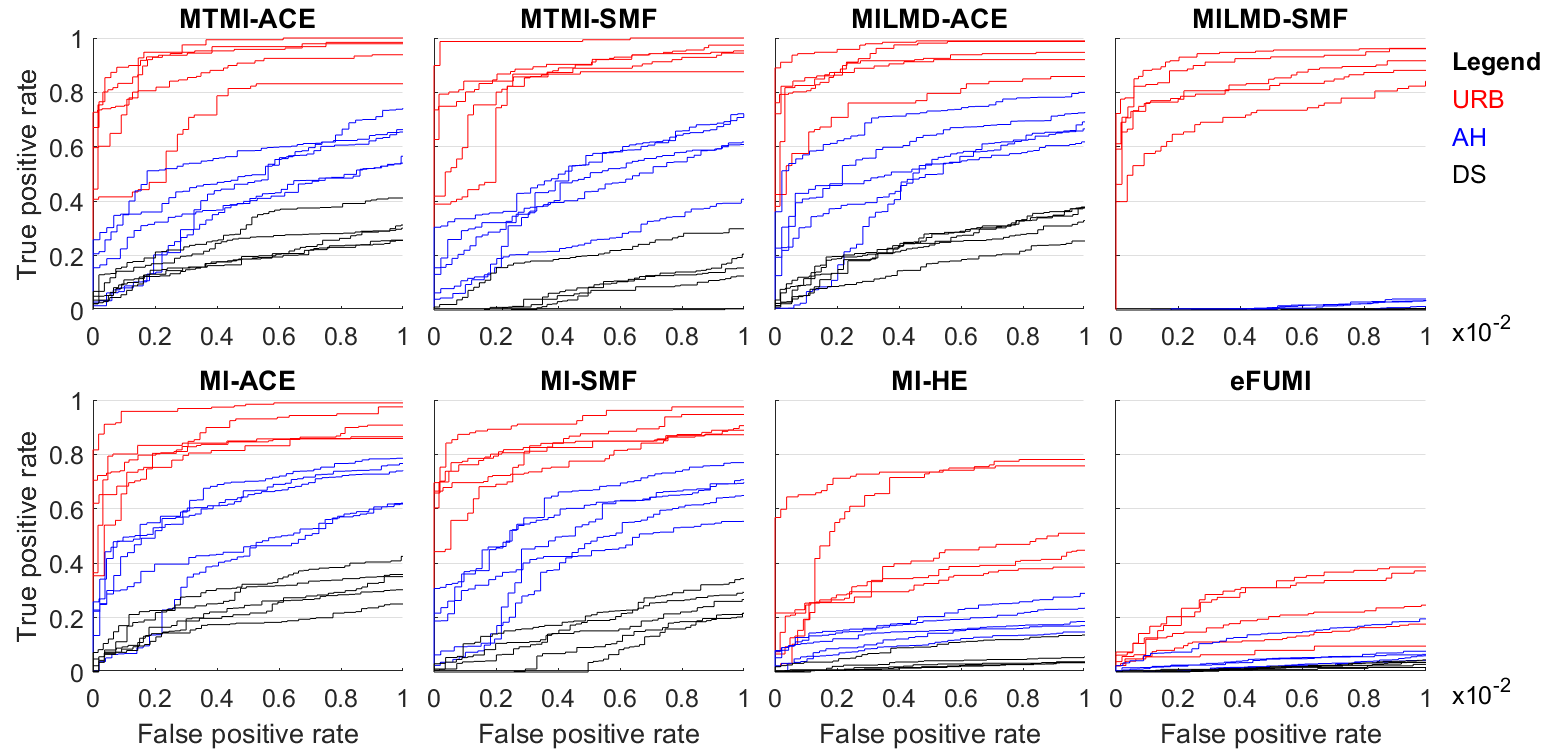}
    \caption{The AVIRIS Santa Barbara dataset ROC curves for three classes (URB: urban, AH: annual herbs, DS: deciduous shrubs) across the 5-Fold iterations. The false-positive rate scale is shown as 1x10$^{-2}$.}
    \label{fig_PFTresults}
\end{figure*}

\subsubsection{Target Detection Results}
Overall, PFTs were detected with high accuracy, considering how much spectral variability is contained in each class (Table \ref{tab:AVIRISresults}; Figure \ref{fig_PFTresults}). Each of these classes contains multiple plant species that were spread across the 10 flight lines, which covered approximately 12,980 km\textsuperscript{2}. This dataset includes spectral variability that is inherent in all plant spectral datasets, which is caused by differences in properties such as plant structure, biochemistry, and water status. However, additional spectral variability is added due to the large spatial extent of this dataset. Classes that are spectrally more homogenous due to the plant species sharing similar plant properties (e.g., AH, EBT) performed the best. Classes with more spectral variability due to significant differences in species (e.g., EBS, ENS) or PFTs that are just difficult to map due to open canopies (e.g., ENT) had lower performance. For example, ENS and ENT classes can have significant overlap between classes, but MTMI-ACE/MTMI-SMF algorithms were able to distinguish the positive and negative bags from each other and yield appropriate target signatures. While on the other hand, the URB class was easily detected by most algorithms because it is so different from the negative bags, which in this case, were mostly vegetation. 

The ability to determine multiple targets for target detection gave MTMI-ACE and MTMI-SMF a boost in performance compared to the single target detection algorithms (MI-ACE and MI-SMF). The exception to this is the DS and EBT classes, which had better NAUC results using MI-ACE and MI-SMF. These results demonstrate that even fewer targets could have been returned using MTMI-ACE and MTMI-SMF for these classes. In general, MTMI performs consistently well across all PFT classes, while other algorithms had more variability results across the classes. 

The MILMD-ACE algorithm, another multiple target multiple instance algorithms, also performed well when detecting PFT targets. Average NAUC values between MTMI-ACE, MTMI-SMF, and MILMD-ACE were often very comparable with the standard deviations. For example, with the RS class, the MILMD algorithm performed better based on average NAUC, but once the standard deviation was accounted for, the results were not different. 

The MTMI algorithms outperformed MI-HE and eFUMI. The difficulty with this dataset is knowing how many targets are necessary to capture the spectral variability for high detection accuracy. In these algorithms, the user specifies how many targets or background targets are present in the dataset. That exact number is returned, which may not be the optimal number of targets. The benefit of the MTMI algorithm is that a max number of targets is entered by the user, but the algorithm determines the appropriate quantity for the class using the $\alpha$ parameter. 

\section{Conclusion}
In this work, the MTMI-ACE and MTMI-SMF algorithms for MIL problems are proposed and investigated. Both algorithms can learn multiple discriminative target concepts from ambiguously labeled data. Comprehensive experiments show that the proposed MTMI-ACE and MTMI-SMF algorithms are effective in learning discriminative target concepts. These two algorithms achieved superior performance over other state-of-the-art MIL algorithms in several experiments that tested different target detection scenarios.

Additionally, MTMI-ACE and MTMI-SMF present a few advantages over comparison algorithms. First, the appropriate number of target signatures for a target’s detection is returned, reducing the need for a user's knowledge of the target’s spectral variability. Second, the MTMI-ACE and MTMI-SMF algorithms do not require that more than one target needs to be present in each positive bag, increasing the number of applications. Lastly, these algorithms efficiently determine target signatures compared to other sampling algorithms. Although this paper focuses on hyperspectral target detection, the MTMI-ACE and MTMI-SMF algorithms are a general MIL framework that could be applied to any problems containing mixed and ambiguously labeled training data.

\section*{Acknowledgments}

This project was funded by the Army Research Office with grant number W911NF-17-1-0213 to support the US Army RDECOM CERDEC NVESD. The views and conclusions contained in this document are those of the authors and should not be interpreted as representing the official policies either expressed or implied of the Army Research office, Army Research Laoratory, or the US Government. This project was supported by a DARPA Advanced Plant Technologies grant titled \textit{SENTINEL: SENsing Threats In Natural Environments using Ligand-receptor modules}. The project was also partially supported by the Harris Corporation.


\bibliographystyle{IEEEtran}
\bibliography{./bibtex/bib/references}
%





%
\appendices
\onecolumn
\section{Optimization Update Equation Derivation}

\label{sec:appendixA}

The objective function is written in Equation \eqref{MTMIACEObjFun2}. In Equation \eqref{MTMIACEModifiedobjfun3} the detection statistic function, $D$, is expanded out for the ACE statistic showing the whitened data and inner product. The derivation in this Appendix may be done using the SMF statistic following the same format except $\hat{\hat{\mathbf{x}}}$ would be replaced with $\hat{\mathbf{x}}$.
\begin{figure*}[!h]
    \begin{equation}
        \underset{\textbf{S}}{\max} \quad \frac{1}{N^{+}} \sum_{j:L_{j} = 1} \underset{s_{k}\in \mathbf{S}}\max (D(\mathbf{x}_{j,k}^{*}, \mathbf{s}_{k})) - \frac{1}{N^{-}} \sum_{j:L_{j} = 0} \frac{1}{N_{j}^{-}} \sum_{x_{i} \in B_{j}^{-}} D(\mathbf{x}_{i}, \mathbf{s}_{k}) - \frac{\alpha}{{K \choose 2}}\sum_{k,l, l \neq k} D(\mathbf{s}_{k}, \mathbf{s}_{l}) \quad \text{s.t.} \quad D(\mathbf{s}_{k}, \mathbf{s}_{k}) = 1
        \label{MTMIACEObjFun2}
    \end{equation}
\end{figure*}
\begin{figure*}[!h]
    \begin{equation}
        \underset{\textbf{S}}{\max} \quad \frac{1}{N^{+}} \sum_{j:L_{j} = 1} \underset{\hat{\hat{s}}_{k}\in \mathbf{S}}\max \big(\hat{\hat{\mathbf{x}}}_{j,k}^{*T} \hat{\hat{\mathbf{s}}}_{k}\big) - \frac{1}{N^{-}} \sum_{j:L_{j} = 0} \frac{1}{N_{j}^{-}} \sum_{\hat{\hat{x}}_{i} \in B_{j}^{-}}  \big(\hat{\hat{\mathbf{x}}}_{i}^{T} \hat{\hat{\mathbf{s}}}_{k}\big) - \frac{\alpha}{{K \choose 2}}\sum_{k,l, l \neq k} \big(\hat{\hat{\mathbf{s}}}_{k}^{T} \hat{\hat{\mathbf{s}}}_{l}\big) \quad \text{s.t.} \quad \hat{\hat{\mathbf{s}}}_{k}^{T} \hat{\hat{\mathbf{s}}}_{k} = 1
        \label{MTMIACEModifiedobjfun3}
    \end{equation}
\end{figure*}
\FloatBarrier
The optimal update equation for each target signature, $\hat{\hat{\mathbf{s}}}_{k}$, can be solved for using the associated Lagrangian, written in Equation \eqref{MTMIACELagrange}.
\begin{figure*}[!h]
    \begin{equation}
        \mathcal{L} = \frac{1}{N^{+}} \sum_{j:L_{j} = 1} \underset{\hat{\hat{s}}_{k}\in \mathbf{S}}\max \big(\hat{\hat{\mathbf{x}}}_{j,k}^{*T} \hat{\hat{\mathbf{s}}}_{k}\big) - \frac{1}{N^{-}} \sum_{j:L_{j} = 0} \frac{1}{N_{j}^{-}} \sum_{\hat{\hat{x}}_{i} \in B_{j}^{-}} \big(\hat{\hat{\mathbf{x}}}_{i}^{T} \hat{\hat{\mathbf{s}}}_{k}\big) - \frac{\alpha}{{K \choose 2}}\sum_{k,l, l \neq k} \big(\hat{\hat{\mathbf{s}}}_{k}^{T} \hat{\hat{\mathbf{s}}}_{l}\big) - \lambda \big(\hat{\hat{\mathbf{s}}}_{k}^{T} \hat{\hat{\mathbf{s}}}_{k} - 1\big)
        \label{MTMIACELagrange}
    \end{equation}
\end{figure*}

\FloatBarrier
The derivative of the Lagrangian with respect to the target signature is taken and shown in Equation \eqref{MTMIACELagrangeDerive}. Here the max operation on the first term is expanded out using an indicator function.
\FloatBarrier
\begin{figure*}[!h]
    \begin{equation}
        \frac{\partial \mathcal{L}}{\partial \hat{\hat{\mathbf{s}}}_{k}} = \frac{1}{N_{k}^{+}} \sum_{j:L_{j} = 1} \mathbb{1}_{j,k}^{+} \hat{\hat{\mathbf{x}}}^{*}_{j,k} - \frac{1}{N^{-}} \sum_{j:L_{j} = 0} \frac{1}{N_{j}^{-}} \sum_{\hat{\hat{x}}_{i} \in B_{j}^{-}} \hat{\hat{\mathbf{x}}}_{i} - \frac{\alpha}{(K-1)} \sum_{l, l \neq k} \hat{\hat{\mathbf{s}}}_{l} - 2\lambda \hat{\hat{\mathbf{s}}}_{k} \ ,
        \label{MTMIACELagrangeDerive}
    \end{equation}
\end{figure*}
The positive bag indicator for target signature $\hat{\hat{\mathbf{s}}}_{k}$ is defined as  
\begin{equation}
    \mathbb{1}_{j,k}^{+} =
    \begin{cases}
        1 & \text{if} \quad  \hat{\hat{\mathbf{x}}}^{*T}_{j,k} \hat{\hat{\mathbf{s}}}_{k} \ > \ \hat{\hat{\mathbf{x}}}^{*T}_{j,l} \hat{\hat{\mathbf{s}}}_{l}, \quad \forall l \neq k \\
        0 & \text{otherwise}
    \end{cases}
    \label{MTMIACEOptimizationBagIdentifier2}
\end{equation}

Then solving for the target signature the update equation and the Lagrangian multiplier is solved for in equations \eqref{MTMIACELagrangeDerive2} and \eqref{MTMIACELagrangeDerive3}.
\begin{figure*}[!h]
    \begin{equation}
        \hat{\hat{\mathbf{s}}}_{k} = \frac{1}{2\lambda} \Bigg( \frac{1}{N_{k}^{+}} \sum_{j:L_{j} = 1} \mathbb{1}_{j,k}^{+} \hat{\hat{\mathbf{x}}}^{*}_{j,k} - \frac{1}{N^{-}} \sum_{j:L_{j} = 0} \frac{1}{N_{j}^{-}} \sum_{\hat{\hat{x}}_{i} \in B_{j}^{-}} \hat{\hat{\mathbf{x}}}_{i} - \frac{\alpha}{(K-1)} \sum_{l, l \neq k} \hat{\hat{\mathbf{s}}}_{l} \Bigg)
        \label{MTMIACELagrangeDerive2}
    \end{equation}
\end{figure*}

\begin{figure*}[!h]
    \begin{equation}
        \lambda = \frac{\vert \vert \hat{\hat{\mathbf{t}}} \vert \vert}{2}
        \label{MTMIACELagrangeDerive3}
    \end{equation}
\end{figure*}

\FloatBarrier
Finally, the update equation for the $k^{th}$ target signature with the lagrangian multiplier is shown in Equation \eqref{MTMIACEUpdate2}.
\begin{figure*}[!h]
    \begin{equation}
        \hat{\hat{\mathbf{s}}}_{k} = \frac{\hat{\hat{\mathbf{t}}}}{\left \| \hat{\hat{\mathbf{t}}} \right \|} \textup{ where } \hat{\hat{\mathbf{t}}} = \frac{1}{N_{k}^{+}} \sum_{j:L_{j} = 1} \mathbb{1}_{j,k}^{+} \hat{\hat{\mathbf{x}}}^{*}_{j,k} - \frac{1}{N^{-}} \sum_{j:L_{j} = 0} \frac{1}{N_{j}^{-}} \sum_{\hat{\hat{x}}_{i} \in B_{j}^{-}} \hat{\hat{\mathbf{x}}}_{i} - \frac{\alpha}{(K-1)} \sum_{l, l \neq k} \hat{\hat{\mathbf{s}}}_{l}
        \label{MTMIACEUpdate2}
    \end{equation}
\end{figure*}
\FloatBarrier
\twocolumn
\section{Parameters for Experiments}

\subsection{Simulated Dataset}

\begin{table}[h]
    \setlength\extrarowheight{2pt}
    \centering
    \begin{tabular}{|c c|}
        \hline
         Algorithm & Parameters\\
         \hline
         MTMI-ACE & \textit{K} = 4, $\alpha$ = 1\\
         MTMI-SMF & \textit{K} = 4, $\alpha$ = 1\\
         MILMD-ACE & \textit{K} = 2, $\alpha$ = 0.5\\
         MILMD-SMF & \textit{K} = 2, $\alpha$ = 0.5\\
         MI-ACE  & \textit{K} = 1, $\alpha$ = 0\\
         MI-SMF  & \textit{K} = 1, $\alpha$ = 0\\
         MI-HE   & \textit{T} = 2, \textit{M} = 3, $\rho$ = 0.8, \textit{b} = 5, $\beta$ = 5, $\lambda$ = $1x10^{-3}$\\
         eFUMI   & \textit{M} = 3, $\alpha$ = 1.2, $\beta$ = 60, $\Gamma$ = 10, \textit{u} = 0.05  \\
         \hline
    \end{tabular}
    \label{tab:SimParams_Sim}
\end{table}

\subsection{MUUFL Gulfport Dataset}

\begin{table}[h]
    \setlength\extrarowheight{2pt}
    \centering
    \begin{tabular}{|c c|}
        \hline
         Algorithm & Parameters\\
         \hline
         MTMI-ACE & \textit{K} = 2, $\alpha$ = 0.1\\
         MTMI-SMF & \textit{K} = 2, $\alpha$ = 0.1\\
         MILMD-ACE &  \textit{K} = 2, $\alpha$ = 0.1\\
         MILMD-SMF & \textit{K} = 2, $\alpha$ = 0.1\\
         MI-ACE  & \textit{K} = 1, $\alpha$ = 0\\
         MI-SMF  & \textit{K} = 1, $\alpha$ = 0\\
         MI-HE   & \textit{T} = 1, \textit{M} = 9, $\rho$ = 0.3, \textit{b} = 5, $\beta$ = 1, $\lambda$ = $5x10^{-3}$\\
         eFUMI   & \textit{M} = 7, $\alpha$ = 2, $\beta$ = 60, $\Gamma$ = 5, \textit{u} = 0.05\\
         \hline
    \end{tabular}
    \label{tab:SimParams_MUUFL}
\end{table}

\subsection{AVIRIS Santa Barbara Dataset}

\begin{table}[h]
    \setlength\extrarowheight{2pt}
    \centering
    \begin{tabular}{|c c|}
        \hline
         Algorithm & Parameters\\
         \hline
         MTMI-ACE & \textit{K} = 10, $\alpha$ = 0.05 \\
         MTMI-SMF & \textit{K} = 10, $\alpha$ = 0.05\\
         MILMD-ACE &  \textit{K} = 10, $\alpha$ = 0.05\\
         MILMD-SMF & \textit{K} = 10, $\alpha$ = 0.05\\
         MI-ACE  & \textit{K} = 1, $\alpha$ = 0\\
         MI-SMF  & \textit{K} = 1, $\alpha$ = 0\\
         MI-HE   & \textit{T} = 10, \textit{M} = 20, $\rho$ = 0.8, \textit{b} = 5, $\beta$ = 1, $\lambda$ = $5x10^{-3}$\\
         eFUMI   & \textit{M} = 20, $\alpha$ = 2, $\beta$ = 60, $\Gamma$ = 5, \textit{u} = 0.05\\
         \hline
    \end{tabular}
    \label{tab:SimParams_AVIRIS}
\end{table}


\ifCLASSOPTIONcaptionsoff
  \newpage
\fi

\end{document}